\documentstyle[11pt,fleqn,cite]{article}   
\catcode`\@=11
\@addtoreset{equation}{section}
\def\theequation{\arabic{section}.\arabic{equation}}
\def\appendix{\renewcommand{\thesection}{\Alph{section}}\setcounter{section}{0}
              \renewcommand{\theequation}
            {\mbox{\Alph{section}.\arabic{equation}}}\setcounter{equation}{0}}
\def\maketitle{\thispagestyle{empty}\setcounter{page}0\newpage
                \renewcommand{\thefootnote}{\arabic{footnote}}
                  \setcounter{footnote}0}
\renewcommand{\thanks}[1]{\renewcommand{\thefootnote}{\fnsymbol{footnote}}
               \footnote{#1}\renewcommand{\thefootnote}{\arabic{footnote}}}
\newcommand{\preprint}[1]{\hfill{\sl preprint - #1}\par\bigskip\par\rm}
\renewcommand{\title}[1]{\begin{center}\Large\bf #1\end{center}\rm\par\bigskip}
\renewcommand{\author}[1]{\begin{center}\Large #1\end{center}}
\newcommand{\address}[1]{\begin{center}\large #1\end{center}}

\def\dinfn{\smallskip Dipartimento di Fisica, Universit\`a di Trento\\ 
                           and Istituto Nazionale di Fisica Nucleare,\\
                                   Gruppo Collegato di Trento, Italia}

\def\Idinfn{\address{\dinfn}}
\newcommand{\email}[1]{e-mail: \sl #1@science.unitn.it\rm}
\newcommand{\femail}[1]{\thanks{\email{#1}}}
\newcommand{\pacs}[1]{\smallskip\noindent{\sl PACS numbers:
                       \hspace{0.3cm}#1}\par\bigskip\rm}
\def\babs{\hrule\par\begin{description}\item{Abstract: }\it} 
\def\eabs{\par\end{description}\hrule\par\medskip\rm}
\renewcommand{\date}[1]{\par\bigskip\par\sl\hfill #1\par\medskip\par\rm}
\newcommand{\ack}[1]{\par\section*{Acknowledgments} #1} 
\newcommand{\s}[1]{\section{#1}}

\renewcommand{\vec}[1]{{\bf #1}}       %%%  vectors in bold
\def\M{{\cal M}}                       %%%  calligraphic M 
\newcommand{\ca}[1]{{\cal #1}}         %%%  calligraphic
\def\hs{\qquad}               %%%  horizontal space
\def\nn{\nonumber}            %%%  no number for eqnarray
\def\beq{\begin{eqnarray}}    %%%  begequation/eqnarray
\def\eeq{\end{eqnarray}}      %%%  endequation/eqnarray
\def\at{\left(}               %%%  open (
\def\aq{\left[}               %%%  open [
\def\ct{\right)}              %%%  close )
\def\cq{\right]}              %%%  close ]
\def\R{{\hbox{{\rm I}\kern-.2em\hbox{\rm R}}}}   %%% real numbers
\def\H{{\hbox{{\rm I}\kern-.2em\hbox{\rm H}}}}   %%% Hilbert space
\def\N{{\hbox{{\rm I}\kern-.2em\hbox{\rm N}}}}   %%% natural numbers
\def\C{{\ \hbox{{\rm I}\kern-.6em\hbox{\bf C}}}} %%% complex numbers
\def\Z{{\hbox{{\rm Z}\kern-.4em\hbox{\rm Z}}}}   %%% integers numbers
\def\ii{\infty}                                  %%% infinit
\def\X{\times\,}                                 %%% times
\newcommand{\fr}[2]{\mbox{$\frac{#1}{#2}$}}      %%% small fraction
\def\Tr{\mathop{\rm Tr}\nolimits}                  %%% Trace
\def\lap{\Delta}                                   %%% Laplacian
\def\be{\beta}
\def\ga{\gamma}

\def\ep{\varepsilon}
\def\ze{\zeta}

\def\ka{\kappa}

\def\si{\sigma}

\def\ph{\varphi}

\def\te{\vartheta}

\def\De{\Delta}
\def\La{\Lambda}
\def\Si{\Sigma}
\def\Om{\Omega}

\begin{document}
%\tableofcontents       %%%%%%   index of section

\preprint{UTF 407, gr-qc/9710118}
\title{Thermodynamics of scalar fields in Kerr's geometry}

\author{Guido Cognola\femail{cognola}}
\Idinfn

\babs
The one-loop contributions to the entropy for a massive scalar field in a 
Kerr black hole are investigated using an approximation of the metric, 
which, after a conformal transformation, permits to work in a 
Rindler-like spacetime. 
Of course, as for the Schwarzschild case, the entropy is divergent
in the proximity of the event horizon.
\eabs


\s{Introduction}

As has been recently stressed in a series of papers
(see for example \cite{frol96-54-2711} and references cited therein),
the Bekenstein-Hawking entropy 
\cite{beke73-7-2333,hawk75-43-199,gibb77-15-2752}, that
for a stationary black hole can be computed by several methods,
which lead to the celebrated result $A/4$
(tree-level contribution) (see for example \cite{iyer95-52-4430}), 
has a thermodynamical origin in the sense
that it can be defined by the response of the free energy 
of the black hole to the change of the equilibrium temperature. 
Such a temperature depends on the parameters of the black hole and may 
be determined by requiring the smoothness of the related Euclidean solution 
\cite{gibb77-15-2752} (on-shell computation).

The situation is completely different if one tries to investigate 
the entropy within a statistical-mechanical approach,
that is by counting the quantum states of the black hole. 
In fact in this case one is forced to work at an arbitrary temperature,
which is not the equilibrium one (off-shell computation).  
The first computation of this kind has been appeared in 
the 't~Hooft seminal paper \cite{thoo85-256-727}, 
where the black hole degrees of freedom have 
been identified with the ones of a quantum gas of scalar particles 
propagating outside the horizon at a given temperature $T$. 
After that, the 't~Hooft brick wall model has been extended to other
geometries \cite{ho97-14-2617} and to other fields \cite{cogn97u-402} and
also a lot of different methods have been proposed in order to
compute the entropy of fields in the black hole geometry 
(see for example Refs.~\cite{frol96-54-2711,cogn95-12-1927,ghos97-78-1858}
and references cited therein).

As is well known, independently of the method of computation used,
the statistical-mechanical quantities were found to be divergent. 
These divergences are not totally unexpected. 
In fact their physical origin may be derived by the equivalence principle,
which  implies that a system in thermal equilibrium has a local Tolman 
temperature given by $T(x)=T_\ii/{\sqrt {|g_{00}(x)|}}$, 
$T_\ii$ being the temperature measured by the observer 
at the spatial infinity. 
The leading term in the high temperature expansion for the free energy 
of a massless quantum gas in a 4-dimensional static space-time 
is proportional to the integral of $T^4(x)$ over the space variable
and this is divergent since the metric has non integrable 
singularities on the horizon. 
The nature of these divergences depends on the poles and zeros of the metric.
For example, for extreme black holes, where  $g_{00}$ has higher order 
zeros, the divergences are much more severe than those which appear 
in the non extremal case (see for example \cite{cogn95-52-4548}). 

In the present paper we shall compute the one-loop contribution to the 
entropy of a Kerr black hole due to a massive scalar field.
Such a problem has been already studied in Ref.~\cite{ho97-14-2617} using
a semiclassical approach and, for the massless case,
in Ref.~\cite{mann96-54-3932} using the method 
of ``blunt conical singularity''. 
Here we shall compute the free energy through the Euclidean path integral 
and using heat kernel and  $\zeta$-function regularisation methods.
In order to perform explicit calculations, 
we have to use an approximation for the Kerr metric, 
valid in the proximity of the horizon, which, after a conformal transformation,
permits to work in a Rindler-like manifold.

At first sight, the present approach could be confused with the one 
proposed in Ref.~\cite{mann96-54-3932}. In fact, in that paper
the authors use path integral and heat kernel techniques too and also
an approximation for the metric very similar to our, but the use of it
and its interpretation is different. 
Moreover in that paper, the singularity,
which naturally appears when arbitrary temperature is considered,
is regularised by means of a family of smooth manifolds, 
while here we prefer to work in the original manifold with the 
conical singularity, since heat kernel and $\zeta$-function
are well known in such kind of spaces.  

In the paper we shall use conformal transformation techniques, 
which are resumed in Sec.~\ref{S:conf} and $\zeta$-function regularisation
on Rindler-like manifold, which we recall in Sec.~\ref{S:Rind}.
In Sec.~\ref{S:Kerr} we propose an approximation of the Kerr metric,
valid in the proximity of the horizon and finally, in 
Sec.~\ref{S:TSFKerr}, we compute the entropy  for a massive scalar field
in the Kerr geometry and conclude with some comments in 
Sec.~\ref{S:Conclusion}.

\s{Conformal transformations} 
\label{S:conf}

Conformal transformation techniques have been used by many authors in 
order to transform the original static manifold in an ultrastatic one
(optical manifold) 
\cite{gibb78-358-467,dowk78-11-895,page82-25-1499,brow85-31-2514,
gusy87-46-1097,dowk88-38-3327,dowk89-327-267}.
In the context of black holes, they have been used in 
Refs.~\cite{barv95-51-1741,cogn95-12-1927,zerb96-54-2699}
and in Rindler space-times in 
Refs.~\cite{byts96-458-267,iell96-54-7459,more97-55-3552}.  

In the following we do not work in the optical manifold, but nevertheless
we shall use conformal transformations in order to simplify the metric. 
A conformal transformation does not modify the temperature 
dependent part of the free energy
of the system and so one can compute all thermodynamical quantities 
in the transformed metric and then simply write them in the original one.

To start with, we consider a scalar field
on a 4-dimensional static space-time with metric 
$g_{\mu\nu}(\vec x)$ and signature $\{-+++\}$ ($\mu,\nu=0,...,3$). 
The one-loop partition function at temperature $T=1/\be$ is given by 
(as usual we perform the Wick rotation $x_0=-i\tau$
and assume the field to be periodic in the $\tau$ variable, with period $\be$)
\begin{equation}
Z_\be=\int d[\phi]\,
\exp\left(-\frac12\int\phi L_4\phi\:d^4x\right)
\:,\end{equation}
where $\phi$ is a scalar density of weight $-1/2$ and
$L_4$ is a Laplace-like operator on the 4-dimensional manifold.
It has the form
\begin{eqnarray}
L_4=-\Delta_4+m^2+\xi R
\:.\end{eqnarray}
Here $\Delta_4$ is the Laplace-Beltrami operator in the metric $g$, 
$m$ (the mass) and $\xi$ arbitrary parameters and 
$R$ the scalar curvature of the manifold. 

Now we perform the conformal transformation 
\begin{eqnarray}
\bar{g}_{\mu\nu}(\vec{x})=e^{2\sigma(\vec{x})}g_{\mu\nu}(\vec{x})
\:,\end{eqnarray}
\begin{eqnarray} 
L_\si=\bar L_4=e^{-\sigma}L_4e^{-\sigma}=
-\bar\Delta_4+\frac16\bar R
+e^{-2\sigma}\left[m^2+\at\xi-\fr16\ct R\right]
\:,\label{aconf}\end{eqnarray}
where $\sigma(\vec{x})$ is a suitable function (we shall use over-bar 
symbols for quantities related to the metric $\bar g$). 
The one-loop partition function transforms as 
\begin{eqnarray}
\bar Z_\be=J[g,\bar{g}]\,Z_\be
\:,\end{eqnarray}
where $J[g,\bar{g}]$ is the Jacobian of the conformal transformation.
Such a Jacobian can be explicitly computed \cite{dowk89-327-267},
but for our purposes it is sufficient to know that in a static manifold 
it depends linearly on $\be$. In fact it can be expressed as an integral 
over spacetime and over $s$ of a Seeley-DeWitt coefficient related 
to the field operator $L_{s\si}$ 
\cite{gusy87-46-1097,dowk89-327-267,byts96-266-1}. 
Since the manifold is static, 
the integral over the imaginary time trivially gives a $\be$-factor. 
This means that the Jacobian can be ignored in the computation 
of thermodynamical quantities starting from the free energy, 
which is related to the canonical partition function 
by means of the usual relation
\begin{eqnarray} 
F_\beta=-\frac{1}{\beta}\ln Z_\be
=-\frac{1}{\beta}\left(\ln\bar Z_\be-\ln J[g,\bar g]\right)\:,
\hs\bar F_\beta=-\frac{1}{\beta}\ln\bar Z_\be
\:.\label{FE}\end{eqnarray}
From the latter equation one obtains
\begin{eqnarray}
S_\beta=\beta^2\partial_\beta F_\beta
=\beta^2\partial_\beta \bar F_\beta \:.
\label{entropy}
\end{eqnarray}

The partition function $Z_\be$ can be expressed in terms of the determinant 
of the field operator $L_4$ and the determinant can be usefully defined
by using $\ze$-function \cite{hawk77-55-133}. In this way we get
\begin{eqnarray} 
\ln Z_\be=\frac{1}{2}\zeta_\be'(0|L_4/\mu^2)\:,\hs
\ln\bar Z_\be=\frac{1}{2}\zeta_\be'(0|\bar L_4/\mu^2)\:,
\label{lnZ-Zbar}\end{eqnarray}
where $\mu$ is an arbitrary parameter necessary to adjust the 
dimensions and $\zeta'$ represents the derivative 
with respect to $s$ of the $\ze$-function related to the operator
in the argument.  

\s{Heat kernel and $\ze$-function in Rindler-like spacetimes}
\label{S:Rind}

Here we resume the main results concerning the definition of free energy
in Rindler-like spaces. A detailed analysis in arbitrary dimensions   
has been done in Ref.~\cite{zerb96-54-2699}, where we refer the reader for
more details. 

We call Rindler-like spacetime a manifold of the form $\M={\cal R}\X\M_2$,
with the metric $ds^2=ds^2({\cal R})+ds^2(\M_2)$,
${\cal R}$ being the 2-dimensional Rindler spacetime and $\M_2$  
an arbitrary 2-dimensional smooth manifold. The Euclidean metric reads 
\beq
ds^2=x^2 d\tau^2+dx^2+\ga_{ab}(y)dy^ady^b\:,
\nn\eeq
where $\tau$ ($0\leq\tau\leq\be$) is the imaginary time, 
$x\geq 0$ the radial coordinate and $y$ the coordinates on $\M_2$. 
For an arbitrary $\be$ the manifold $\M$ has the topology of 
$C_{\be}\times\M_2$, $C_{\be}$ being the 2-dimensional cone. 
The Laplacian like operator assumes the form
\beq 
L_4=-\lap_\be+L_2=-\lap_\be-\lap_2+f(y)\:,
\eeq
where $\lap_\be$ and $\lap_2$ are the  Laplace operators on 
$C_\be$ and $\ca M_2$ respectively and, for more generality, 
an arbitrary function on $\ca M_2$ has been added too.

The partition function on such kind of spaces can be written in the form
\cite{zerb96-54-2699}
\beq
F_\be=-\frac{A_0(L_2)I_\be(-1)}{4\be\ep^2}
-\frac{A_1(L_2)I_\be(0)}{2\be}\:\ln\frac{\La^2}{\ep^2}
+\frac{2\pi}{\be}F_{2\pi}\:,
\eeq
where $\ep$ and $\La$ are cutoff parameters
and $A_n(L_2)$ are the spectral coefficients related to the 
Laplace-like operator $L_2$ on $\M_2$, that is
\beq 
\Tr e^{-sL_2}\sim\sum_n\:A_n(L_2)\:s^{n-1}\:.
\eeq
The function $F_{2\pi}$ is the free energy on the smooth manifold, 
that is in the absence of the conical singularity. It may have
infrared (volume) divergences related to the cutoff $\La$, 
but is regular for $\ep\to0$ (the horizon).
Finally, the function $I_\be(s)$ is strictly related 
to the $\zeta$-function of
the Laplacian on the cone and was studied in detail in 
Ref.~\cite{zerb96-54-2699} and in
Ref.~\cite[that was called $G_\be$]{cogn97-42-95}. 
Here we only need its values at $s=0,-1$. 
They read
\beq 
I_\be(0)&=&\frac16\:\at\frac{\be}{2\pi}-\frac{2\pi}{\be}\ct\:,
\\ 
I_\be(-1)&=&\frac1{90}\:\aq\at\frac{2\pi}{\be}\ct^3
+10\frac{2\pi}{\be}-11\frac{\be}{2\pi}\cq\:.
\eeq
As expected, for $\be=2\pi$, $I_{2\pi}(0)=I_{2\pi}(-1)=0$. 
In fact in this case the whoole manifold is smooth and 
one does not have singularities for $x\to0$.

The formula for the entropy now reads
\beq
S_\be=\frac{A_0(L_2)}{90\ep^2}
\:\aq\at\frac{2\pi}{\be}\ct^3
  +5\:\frac{2\pi}{\be}\cq
-\frac{\pi A_1(L_2)}3\:\at\frac{2\pi}{\be}\ct\:
\ln\frac{\La}{\ep}-2\pi F_{2\pi}\:,
\label{FEeven}\eeq

\s{The Kerr's black hole}
\label{S:Kerr}

In this section we consider the Kerr solution of Einstein field equations
and we study an approximation of the metric, which is valid near the horizon. 
It is suitable for the analysis of quantum field 
fluctuations near the black hole.
The Kerr solution describes the gravitational field outside 
a rotating body (with axial symmetry) of mass $M$ and 
angular momentum $J=Ma$ and is believed to
be the unique solution for the description of all (uncharged) 
rotating black holes formed by collapse. 
It is usually written in the stationary form
\beq
ds^2&=&-\frac{\rho^2\De}{\Si^2}\:dt^2+\frac{\rho^2}{\De}\:dr^2
+\rho^2d\te^2
+\frac{\Si^2}{\rho^2}\at d\ph-\frac{2aMr}{\Si^2}
\:dt\ct^2\sin^2\te\:,
\eeq
\beq 
\De&=&r^2-2Mr+a^2=(r-r_+)(r-r_-)\:,
\nn\\ 
\Si^2&=&(r^2+a^2)^2-a^2\De\sin^2\te\:,
\nn\\
\rho^2&=&r^2+a^2\cos^2\te\:.
\nn\eeq
The irremovable singularity of space-time is given by 
$\rho^2=0$ and has the structure of a ring, 
while $r_\pm=M\pm\sqrt{M^2-a^2}$ are singularities of the metric only.
The important parameters which characterise such a solution are given by
(in natural units $G=c=\hbar=k_B=1$)
\beq 
r_H&=&r_+=M+\sqrt{M^2-a^2}\:,
\hs\mbox{(event horizon position),}\nn\\
A_H&=&4\pi(r_H^2+a^2)=8\pi M r_H\:,
\hs\mbox{(area of the horizon),}\nn\\
\ka&=&\frac{r_+-r_-}{4Mr_+}\:,\hs\hs\hs\hs\hs\mbox{(surface gravity).}
\eeq
Another important region is the "ergosphere" where $g_{00}>0$. 
It is given by 
\beq 
r_H\leq R_-<r<R_+\:,\hs
R_\pm=M\pm\sqrt{M^2-a^2\cos^2\te}\:.
\eeq
In such a region, a test particle at a fixed (coordinate) distance $r$ 
from the body must rotate (with respect to the inertial observer at infinity)
with angular velocity $\Om(r)=\frac{2aMr}{\Si^2}$. The value of $\Om(r)$
on the event horizon is identified with the angular velocity of the horizon 
itself $\Om_H$. It reads
\beq 
\Om_H=\Om(r_H)=\frac{2aMr_H}{\Si^2(r_H)}=\frac{a}{2Mr_H}\:.
\eeq

Since we are interested in the physics outside and very near the black hole,
we set 
\beq 
x^2=\frac{4(r-r_+)}{r_+-r_-}
\eeq
and develope the metric for small $x$ by taking into account only the
leading contributions. The result is
\beq
ds_H^2=\rho_H^2\at-\ka^2x^2\:dt^2+dx^2+d\te^2\ct
+\frac{\Si_H^2}{\rho_H^2}\sin^2\te\:d\tilde\ph^2\:,
\label{AKM}\eeq
\beq
\tilde\ph=\ph-\Om_H\:t\nn\:,
\eeq
\beq
\Si_H^2=(r_H^2+a^2)^2=4M^2r_H^2\:,
\hs\hs\rho_H^2=r_H^2+a^2\cos^2\te\:,
\nn\eeq
By the suffix $H$ we indicate the quantities evaluated on the horizon. 
They do not depend on $x$, but may depend on $\te$.
Now, by the conformal transformation 
\beq 
\bar g_{\mu\nu}=\rho_H^{-2} g_{\mu\nu}\:,
\hs\hs\si=-\ln\rho_H\:,
\label{ctK}
\eeq
the line element in Eq.~(\ref{AKM}) assumes the Rindler-like form
\beq
d\bar s_H^2&=&-\ka^2x^2\:dt^2+\:dx^2+d\te^2
+f^2(\te)\sin^2\te\:d\tilde\ph^2\:,\label{TAKM} \\
f^2(\te)&=&\at\frac{r_H^2+a^2}{r_H^2+a^2\cos^2\te}\ct^2
=1+\frac{a^2}{2M^2}\sin^2\te+O\at[a/M]^4\ct\:.
\eeq

To deal with finite temperature field theory,
the imaginary time $\tau=it$ is assumed to be periodic with period $\be$.
For an arbitrary $\be$ the manifold has a conical singularity, which
disappears if $\ka\tau$ has a period equal to $2\pi$. 
To be more precise, the manifold is smooth if the point 
$(\tau,x,\te,\tilde\ph)$ is identified with the point
$(\tau+2\pi/\ka,x,\te,\tilde\ph)$. This means that,
$(\tau,x,\te,\ph)\equiv(\tau+2\pi/\ka,x,\te,\ph-2\pi\Om_H/\ka)$,
in agreement with Gibbons-Hawking prescription \cite{gibb77-15-2752}.
To have an Euclidean section, also the replacement $a\to ia$,
that is $\Om\to i\Om$, has to be performed.
Then, the request that the manifold is smooth fixes the
temperature $T=1/\be$ to the Hawking value \cite{gibb77-15-2752}
\beq 
T_H=\frac{\ka}{2\pi}\:.
\eeq

Now we can use the results of previous sections in order to compute 
the entropy of a scalar field near the horizon of a Kerr black hole.
In has to be stressed that the results will be also valid  for the
Kerr-Newman geometry, since the charge simply modifies the form
of the horizons $r_\pm$.

\s{Thermodynamic of scalar fields in the Kerr's geometry}
\label{S:TSFKerr}

For convenience here we put $\ka=1$. 
The constant parameter $\ka$ will be easily restored 
at the end of calculations.
We consider a minimally coupled, massive scalar field near the horizon of a 
Kerr black hole. The starting point is the D'Alembert operator 
in the original Kerr metric.  
Then we perform the approximation of the metric,
Eq.~(\ref{AKM}), the Wick rotation and the conformal transformation 
(\ref{ctK}) in this way arriving at 
\beq 
\bar L_4=-\lap_\be+L_2\:,\hs
L_2=-\lap_2+\frac{\bar R}{6}-\Om(\te)\:,
\eeq
\beq
\Om(\te)=\rho_H^2\at\frac{R_H}6-m^2\ct\:,
\eeq
according to Eq.~(\ref{aconf}) with $\xi=0$.
In the latter equation, $R_H$ and $\bar R$ are 
the scalar curvatures in the metrics $ds_H^2$ and
$d\bar s_H^2$ respectively, 
$\lap_\be$ is the Laplace operator on the cone $C_\be$, while
$\lap_2$ is a Laplace-like operator 
on the smooth manifold $\M_2$, with metric
\beq 
ds^2(\M_2)=d\te^2+f^2(\te)\sin^2(\te)\:d\tilde\ph^2
\eeq
and curvature $R_2(\te)=\bar R$.
By a straightforward calculation one obtains
\beq 
R_H=\frac{2}{\rho_H^2}-\frac{a^2r_H^2\sin^2\te}{\rho_H^6}\:.
\eeq
Since we only nead the first two spectral coefficients for the operator
$L_2$, the explicit form of $\bar R$ is not strictly  necessary 
and for this reason we do not write down it. 
The important thing is that it depends
only on the coordinates of $\ca M_2$.  

As is well known, 
the first two spectral coefficients are given by
(for a review on spectral geometry see Ref.~\cite{bran90-15-245})  
\beq 
A_0(L_2)=\frac1{4\pi}\int_{0}^{\pi}\int_{0}^{2\pi}
f(\te)\sin\te\:d\te\:d\tilde\ph
=\frac{2M}a\:\arctan\fr{a}{r_H}\:,
\eeq
\beq
A_1(L_2)&=&\frac1{4\pi}\int_{0}^{\pi}\int_{0}^{2\pi}
\Om(\te) f(\te)\sin\te\:d\te\:d\tilde\ph\nn\\
&=&-2m^2Mr_H-\frac{M(r_H^2+3a^2)}{12r_H(r_H^2+a^2)}
-\frac{M(a^2-3r_H^2)}{4ar_H^2}\:\arctan\fr{a}{r_H}\:.
\eeq
Using the latter results in Eq.~(\ref{FEeven}) 
at the equilibrium temperature $\be=2\pi$,
we finally have
\beq 
S_{T_H}&\sim&\frac{A_H}{60\pi\ep^2}\:\frac1{\ka^2\:ar_H}\:
\arctan\fr{a}{r_H}\nn\\
&&\hs+\frac13\aq \frac{m^2A_H}{4\pi}
+\frac{M(r_H^2+3a^2)}{12r_H(r_H^2+a^2)}
+\frac{M(a^2-3r_H^2)}{4ar_H^2}\:\arctan\fr{a}{r_H}
\cq\:\ln\frac{\La}{\ep}
\label{ent}
\eeq
where the constant $\ka$ has been reestablished. Of course, 
in the non rotating case $a\to0$ we obtain the well known 
Schwarzschild result \cite{thoo85-256-727,cogn95-12-1927}.
The leading term of our expression is also compatible with the
one given in Ref.~\cite{mann96-54-3932}.  

We expect the latter equation for the entropy to be valid also 
for the Kerr-Newman black hole, since the charge simply modify the
form of the horizons. This means that in the Kerr-Newman geometry
the quantities $r_H$, $A_H$ and $\ka$ depends also on the charge,
but the form of the entropy is the same.
This result is in contrast with an analog result given in 
Ref.~\cite{mann96-54-3932}. 
In fact in that paper, the contribution to the logarithmic divergence
due to the Kerr-Newman geometry is proportional to the charge and so,
for Schwarzschild and Kerr black holes the logarithmic divergences
are exactly the same.
This is a very surprising result (according to the authors themselves) 
and in our opinion really  strange since, as we said above,
the charge enters only in expressions of $r_\pm$. Then we expect
the final formulae to depend on the charge only throw $r_\pm$.  

\s{Conclusion}
\label{S:Conclusion}

We have derived the one-loop quantum corrections to the entropy of a 
Kerr-Newman black hole due to a massive scalar field, using an 
approximation for the metric valid in a neighbourhood of the horizon. 
As expected, also at the Hawking temperature in the expression for the 
entropy there is a leading divergence, which goes 
as the inverse of the square of the distance from the
horizon and a logarithmic divergence too.
The leading term of Eq.~(\ref{ent}) is compatible with the 
analog expression obtained in Ref.\cite{mann96-54-3932}, 
where a similar approximation for the metric has been used.
As regards the logarithmic contribution, the two expressions
are compatible only in the Schwarzschild case, while for the Kerr-Newman
geometry they are completely different.

As a last comment we observe that the expression for the entropy obtained 
in Ref.~\cite{ho97-14-2617} is more complicated than ours, 
since it contains also a cut-off, which regulates the integration 
in the $\te$ angle and for this reason is really difficult to 
compare that expression with our.

\ack{It is a pleasure to thank L.~Vanzo and S.~Zerbini for
useful discussions.}

\end{document}